\documentclass[prc,preprint,showpacs,preprintnumbers,amsmath,amssymb,superscriptaddress]{revtex4}
\usepackage{graphicx}
\usepackage{dcolumn}
\usepackage{bm}
\usepackage{longtable}
\usepackage{color}
\newcommand{\beqn}{\begin{eqnarray}}
\newcommand{\eeqn}{\end{eqnarray}}
\newcommand{\lb}{\left(}
\newcommand{\rb}{\right)}
\begin{document}

\title{Doublet bands in $^{126}$Cs in the triaxial rotor model coupled with two quasiparticles}
\author{S. Y. Wang}
\affiliation{School of Physics, and MOE Key Laboratory of Heavy Ion
Physics, Peking University, Beijing 100871, China}
\author{S. Q. Zhang}\thanks{e-mail: sqzhang@pku.edu.cn}
\affiliation{School of Physics, and MOE Key Laboratory of Heavy Ion
Physics, Peking University, Beijing 100871, China}
\affiliation{Institute of Theoretical Physics, Chinese Academy of
Science, Beijing 100080, China}
\author{B. Qi}
\affiliation{School of Physics, and MOE Key Laboratory of Heavy Ion
Physics, Peking University, Beijing 100871, China}
\author{J. Meng}\thanks{e-mail: mengj@pku.edu.cn}
\affiliation{School of Physics, and MOE Key Laboratory of Heavy Ion
Physics, Peking University, Beijing 100871, China}
\affiliation{Institute of Theoretical Physics, Chinese Academy of
Science, Beijing 100080, China}
\affiliation{Center of Theoretical Nuclear Physics, National Laboratory of \\
       Heavy Ion Accelerator, Lanzhou 730000,
China}

\date{\today}

\begin{abstract}

The positive parity doublet bands based on the $\pi
h_{11/2}\otimes\nu h_{11/2}$ configuration in $^{126}$Cs have been
investigated in the two quasi-particles coupled with a triaxial
rotor model. The energy spectra $E(I)$, energy staggering parameter
$S(I)=[E(I)-E(I-1)]/2I$, $B(M1)$ and $B(E2)$ values, intraband
$B(M1)/B(E2)$ ratios, $B(M1)_{\textrm{in}}/B(M1)_{\textrm{out}}$
ratios, and orientation of the angular momentum for the rotor as
well as the valence proton and neutron are calculated. After
including the pairing correlation, good agreement has been obtained
between the calculated results and the data available, which
supports the interpretation of this positive parity doublet bands as
chiral bands.
\end{abstract}

 \maketitle

\section{introduction}

Since the original 1997 work of Frauendorf and Meng~\cite{FM97}, the
phenomenon of chiral rotation in atomic nuclei has attracted
significant attention. Lots of experimental and theoretical efforts
have been devoted to search
for~\cite{Starosta01,Koike01,Hartley01,Hecht01,Bark01,Starosta02,Li02,Koike03,
GR03,Vaman04,Joshi04,Wang06a,Tonev06,Petrache06,Grodner06} and to
interpret~\cite{Dimitrov00,Frauendorf01,Peng03a,Peng03b,Koike04,Olbratowski04,Olbratowski06}
this phenomenon.

Following the scenario that the chiral doublet bands are expected to
occur in the mass region where the proton (or neutron) Fermi surface
lies at the bottom of the high-$j$ subshell and the neutron (or
proton) Fermi surface lies at the top of the high-$j$ subshell,
there have been a number of experimental studies searching for
chiral doublet bands in the $A\sim$130 and 100 region. For
$A\sim$130 region, candidate chiral bands have been proposed in
$N=77$ ($^{132}$Cs, $^{134}$La)~\cite{GR03,Bark01}, $N=75$
($^{130}$Cs, $^{132}$La, $^{134}$Pr, $^{136}$Pm,
$^{138}$Eu)~\cite{Starosta01,Hartley01,Hecht01,Starosta02}, $N=73$
($^{128}$Cs, $^{130}$La, $^{132}$Pr)~\cite{Koike01,Grodner06} , and
$N=71$ ($^{126}$Cs)~\cite{Li02,Wang06a} isotones.

The observation of the chirality in $^{126}$Cs, $^{128}$Cs,
$^{130}$Cs, $^{132}$Cs provides a good opportunity to investigate
the dependence of its occurrence on the number of the valence
neutron in high-$j$ shell. In particular, for Cs isotopes, the
proton Fermi levels are supposed to lie in the lower $\pi h_{11/2}$
subshell, and compared with $^{130}$Cs and $^{132}$Cs where the
neutron Fermi levels are supposed to lie in the upper $\nu h_{11/2}$
subshell, the neutron Fermi level in $^{126}$Cs is likely close to
be in the middle of the $\nu h_{11/2}$ subshell. Therefore it is
interesting to examine the mechanism for the occurrence of the
chirality in $^{126}$Cs.

Motivated by these considerations, two quasi-particles coupled with
a triaxial rotor model is employed for the analysis of the positive
parity doublet bands in $^{126}$Cs. The rotational energy spectra
$E(I)$, energy staggering parameter $S(I)=[E(I)-E(I-1)]/2I$, $B(M1)$
and $B(E2)$ values, intraband $B(M1)/B(E2)$ ratios,
$B(M1)_{\textrm{in}}/B(M1)_{\textrm{out}}$ ratios, and orientation
of the angular momentum for the rotor as well as the valence proton
and neutron will be investigated and compared with the data
available. The interpretation of this positive parity doublet bands
as chiral bands will be discussed.

\section{Model}

The particle-rotor model (PRM) has been extensively used in the
investigation of the chiral
rotation~\cite{FM97,Peng03a,Peng03b,Koike03,Koike04} due to the
advantages of the good angular momentum and simple picture. Compared
with the conventional cranking approach, PRM is a quantum mechanical
method which describes the system in the laboratory framework and
yields directly the energy splitting and tunneling between doublet
bands. In Refs.\cite{FM97,Peng03a}, the PRM with one particle and
one hole coupled with a triaxial rotor has been developed and used
in the analysis of chiral bands. The PRM Hamiltonian is written as
 \beqn
    \hat{H}=\hat{H}_{\textrm{core}}+\hat{H}_{\textrm{p}}+\hat{H}_{\textrm{n}},
 \eeqn
 where $\hat{H}_{\textrm{core}}$ represents the Hamiltonian of the
 rotor,
 \beqn
    \hat{H}_{\textrm{core}}=\sum_{\nu=1}^{3}\frac{(I_{\nu}-j_{\textrm{p}\nu}-j_{\textrm{n}\nu})^{2}}{2{\cal J}_{\nu}},
 \eeqn
 ${\cal J}_{\nu} (\nu=1,2,3)$ is the irrotational moment of inertia,
  \beqn
    {\cal J}_{\nu}={\cal J}\sin^{2}(\gamma-\frac{2\pi}{3}\nu) (\nu=1,2,3),
 \eeqn
$H_{\textrm{p}}$ and $H_{\textrm{n}}$ describe the Hamiltonians of
the single proton and neutron outside the rotor which for a single-j
model can be given as,
 \beqn H_{\textrm{p(n)}}=\pm\frac{1}{2}C \left\{ \cos\gamma\lb
j_3^2-\frac{j(j+1)}{3}\rb+\frac{\sin\gamma}{2\sqrt{3}}\lb
j_+^2+j_-^2\rb\right\}
 \eeqn
where the plus sign refers to a particle and the minus to a hole and
the coupling parameter $C$ is proportional to the quadrupole
deformation $\beta$.

Including the pairing by the standard BCS quasiparticle
approximation, the PRM can be generalized to the two quasiparticles
coupled with a triaxial rotor cases. The single-particle energies
$\epsilon_\nu$ obtained by diagonalizing the single proton (neutron)
Hamiltonian in PRM~\cite{FM97,Peng03a} are replaced with the
corresponding quasiparticle energies $E_\nu$,
 \beqn
 E_{\nu \emph{i}}=\sqrt{(\epsilon_{\nu
 \emph{i}}-\lambda_{\emph{i}})^2+\Delta^{2}_{\emph{i}}},
 ~~~~~~~~~~~\emph{i}=\textrm{n, p}
 \eeqn

{where $\varepsilon_{\nu \emph{i}}$
is the single particle energies, $\Delta_{\emph{i}}$ the pairing
gaps and $\lambda_{\emph{i}}$ the Fermi energies. In comparison with
the case excluding pairing, each single-particle matrix element
needs to be multiplied by a pairing factor
$u_{\mu}u_{\nu}+v_{\mu}v_{\nu}$~\cite{Meyer75,Ragnarsson88}, in
which the pairing occupation factor $v_\nu$ of the state $\nu$ is
given by
 \beqn
v^2_\nu=\frac{1}{2}\left[1-\frac{\epsilon_\nu-\lambda}{E_\nu}\right],
 \eeqn
and $ u^2_\nu + v^2_\nu=1$. With inclusion of pairing, the present
model goes beyond the one particle one hole coupled with the
triaxial rotor model, and the configuration of multi-particles
sitting in a high $j$-shell can be simulated by adjusting the Fermi
level. The detailed formalism and numerical calculation will be
published later~\cite{Zhang06}.}

\section{result and discussion}

Using the above two quasiparticles coupled with a triaxial rotor
model, the doublet bands in $^{126}$Cs has been investigated.

In the calculation, the quadrupole deformation
$\varepsilon_{2}=0.244$ and $\gamma=24^\circ$ are taken from
Ref.~\cite{Tajima94}. However, similar as in other PRM
calculations~\cite{Koike04,Fetea05}, in order to achieve better
agreement with the experimental energy spectra a slightly larger
value is taken, i.e., $\varepsilon_{2} = 0.26$. Accordingly, the
parameter $C=0.3$ MeV which corresponds to $\varepsilon_{2}\simeq
0.26$ in the 130 mass region. The moment of inertia ${\cal J}$ = 20
$\hbar^2/$MeV is determined by the slope of the experimental $E$
versus $I$ curve.

Following the empirical formula $\Delta$=12$/\sqrt{A}$, the pairing
gap $\Delta=1.0$ MeV is used for both protons and neutrons. As
mentioned above, as the valence proton in $^{126}$Cs with $Z=55$ and
$N=71$ is supposed to be at the beginning of the $\pi h_{11/2}$
sub-shell, the proton Fermi energy $\lambda_\textrm{p}$ takes value
of $-2.293$ MeV. The neutron Fermi energy $\lambda_\textrm{n}$ takes
value of $0.8$ MeV which lies in the middle of the $\nu h_{11/2}$
sub-shell and simulates the effect of multi-valence neutrons. In the
calculation of the electromagnetic transitions, the empirical
intrinsic qudrupole moment $Q_{0}=3.5$ eb, gyromagnetic ratios
$g_\textrm{R} = Z/A \approx 0.44$, $g_\textrm{p}=1.21$ and
$g_\textrm{n}=-0.21$ have been adopted.

\begin{figure}
\includegraphics[width=12cm, bb=20 200 660 660]{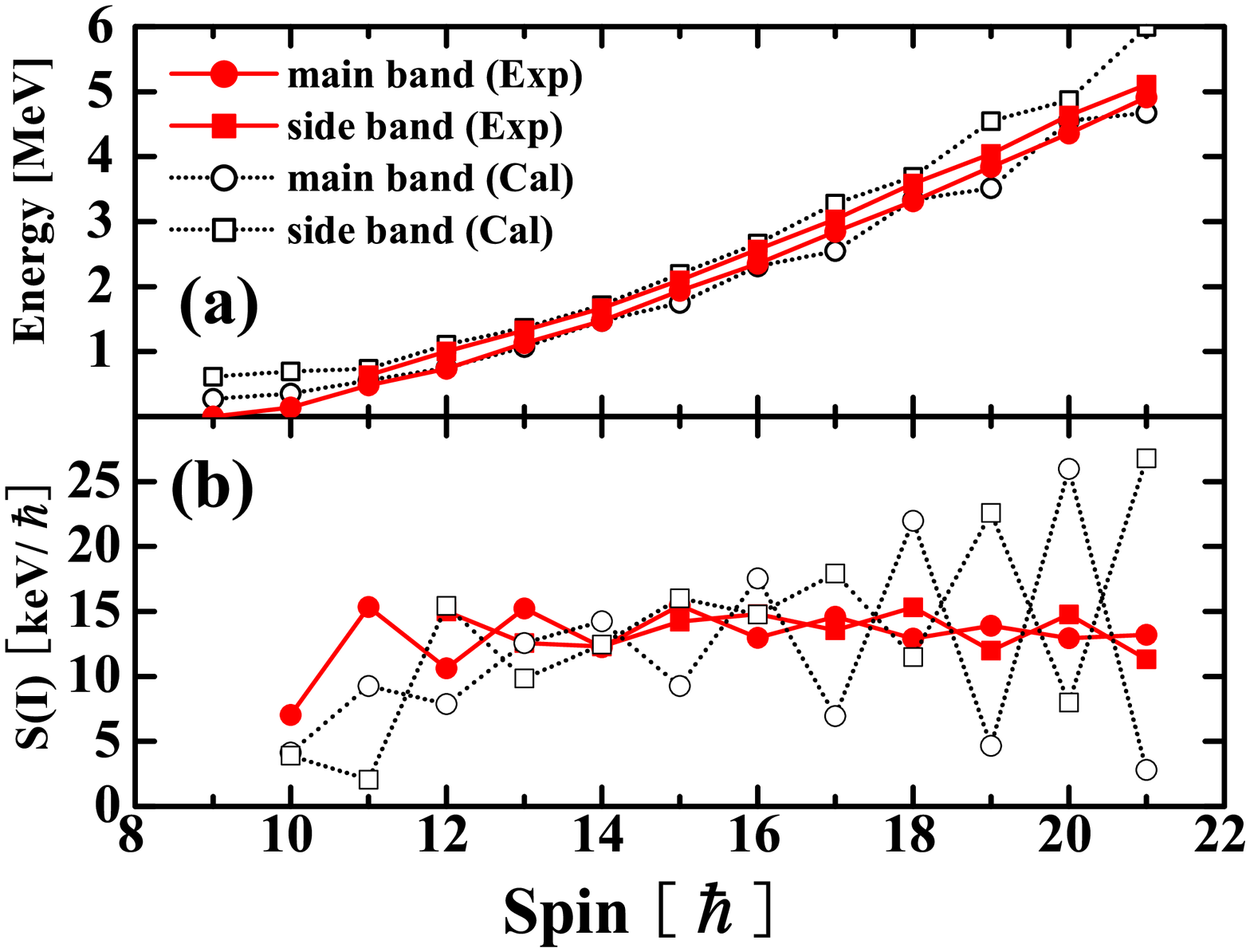}
\caption{\label{fig:Energyandsp} (Color online) The excitation
energy (top panel) and staggering $S(I)=[E(I)-E(I-1)]/2I$ (bottom
panel) as a function of spin for the main and side band in
$^{126}$Cs. The filled (open) symbols connected by solid (dotted)
lines are for the experimental (theoretical) values. The main and
side band are shown by circles and squares, respectively. The
theoretical values in the upper panel are shifted by $-2.65$ MeV in
order to coincide with the experimental energy at $I=14~\hbar$.}
\end{figure}

The calculated energy spectra $E(I)$ and the energy staggering
parameter $S(I)$, defined as $[E(I)-E(I-1)]/2I$, for the doublet
bands in $^{126}$Cs are presented in Fig. \ref{fig:Energyandsp},
together with the corresponding experimental results. The calculated
energy spectra well reproduce the experimental results, and show a
constant energy separation of $\sim$ 200 keV between the two bands
at large spin interval 9$\hbar\leq I\leq$18$\hbar$.  {The energy
staggering parameter $S(I)$ is indicative of the degree of mutual
orthogonality of the three vectors involved. For ideal chiral bands,
$S(I)$ should be spin independent~\cite{Koike02}.} In  Fig.
\ref{fig:Energyandsp}, it can be seen that the theoretical $S(I)$
values reproduce experimental one well. For $I\geq18\hbar$, the
theoretical $S(I)$ values overestimate the amplitude of staggering,
which is due to the large Coriolis effect at high spins. It can be
expected that the attenuation of Coriolis effect may reduce the
staggering discrepancy of $S(I)$.

\begin{figure}
\includegraphics[width=12cm]{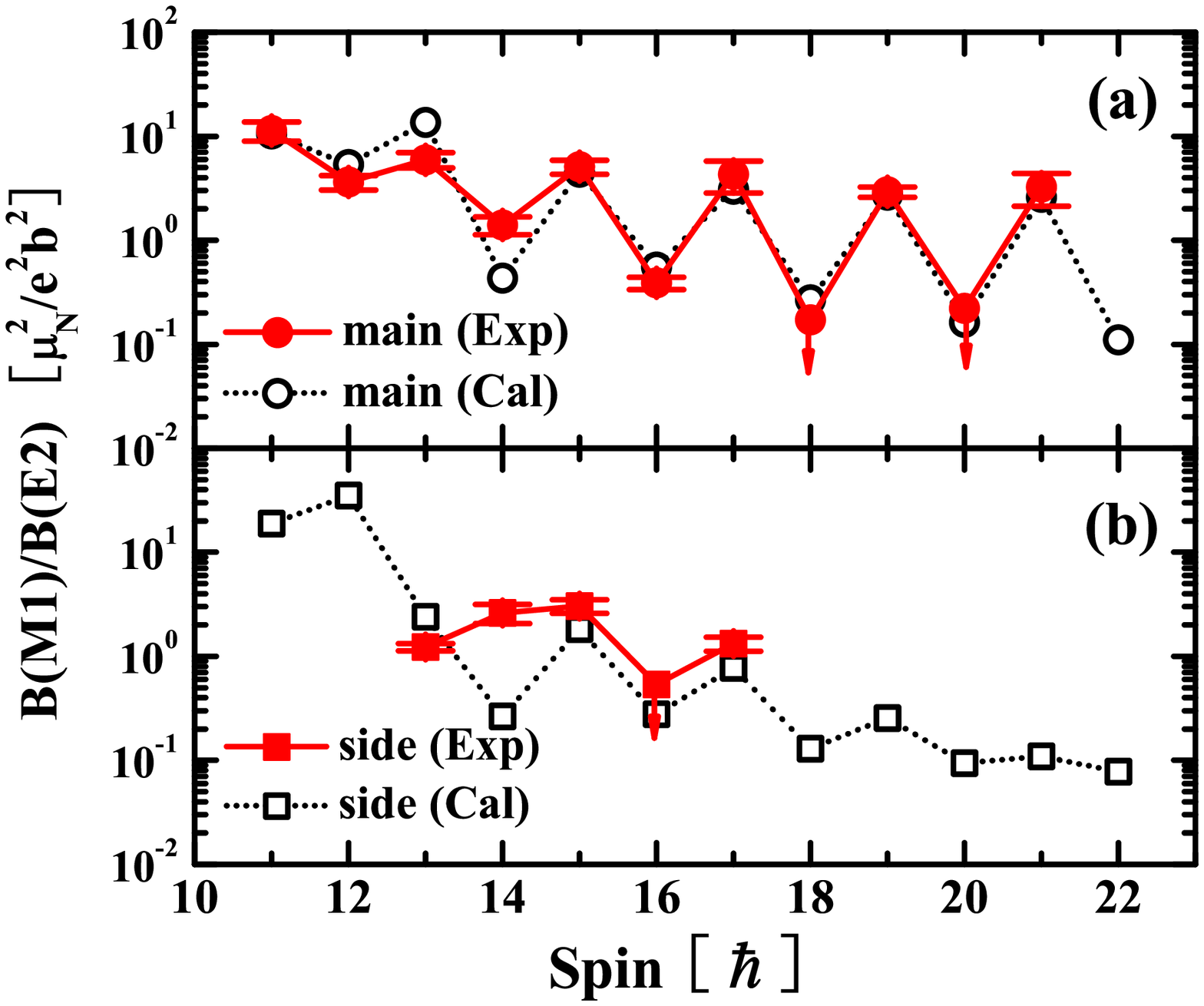}
\caption{\label{fig:Bm1e2} (Color online) Comparisons between the
calculated $B(M1)/B(E2)$ and data available for the main and side
band in $^{126}$Cs. }
\end{figure}

\begin{figure}
\includegraphics[width=12cm]{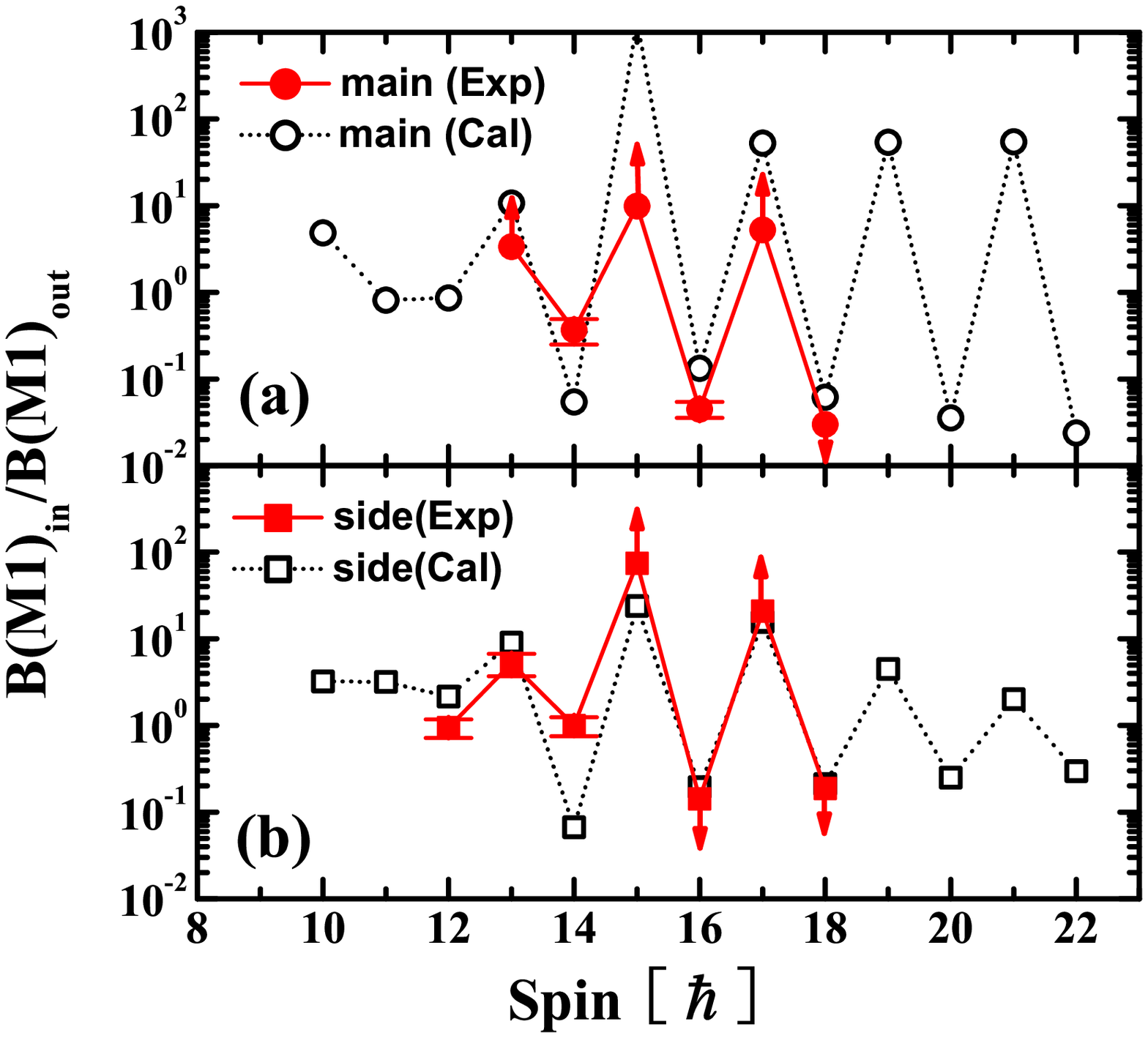}
\caption{\label{fig:Bm1inout} (Color online) Comparisons between the
calculated $B(M1)_{in}/B(M1)_{out}$ and data available for the main
and side band in $^{126}$Cs.}
\end{figure}

It has been suggested in ref.~\cite{Koike02} that the observed
staggering of the ratios $B(M1)/B(E2)$ and
$B(M1)_{\textrm{in}}/B(M1)_{\textrm{out}}$ (the subscript $in$ and
$out$ represent respectively the intraband and interband) can be
understood as signatures of the chiral bands. The calculated
$B(M1)/B(E2)$ and $B(M1)_{\textrm{in}}/B(M1)_{\textrm{out}}$ values
together with their corresponding experimental results for the
doublet bands in $^{126}$Cs are presented in the Fig.
\ref{fig:Bm1e2} and \ref{fig:Bm1inout}, respectively. It can be seen
that the agreement for the $B(M1)/B(E2)$ ratios at the whole spin
region is excellent. The magnitude, staggering and the decreasing
trend of the ratios with spin are reproduced quite well. Furthermore
the experimental staggering phase is exactly reproduced in the
calculation, i.e., the value at odd spin is larger than the one at
even spin. Similarly, the calculated
$B(M1)_{\textrm{in}}/B(M1)_{\textrm{out}}$ ratios also reproduce the
experimental magnitude, staggering and the trend pattern quite well.

In comparison with Ref.~\cite{Peng03a}, one can draw the conclusion
that after including the pairing correlation, not only the energy
spectra but also the electromagnetic transition ratios can be well
reproduced by the two quasi-particles coupled with a triaxial rotor
model.

\begin{figure}
\includegraphics[width=12cm]{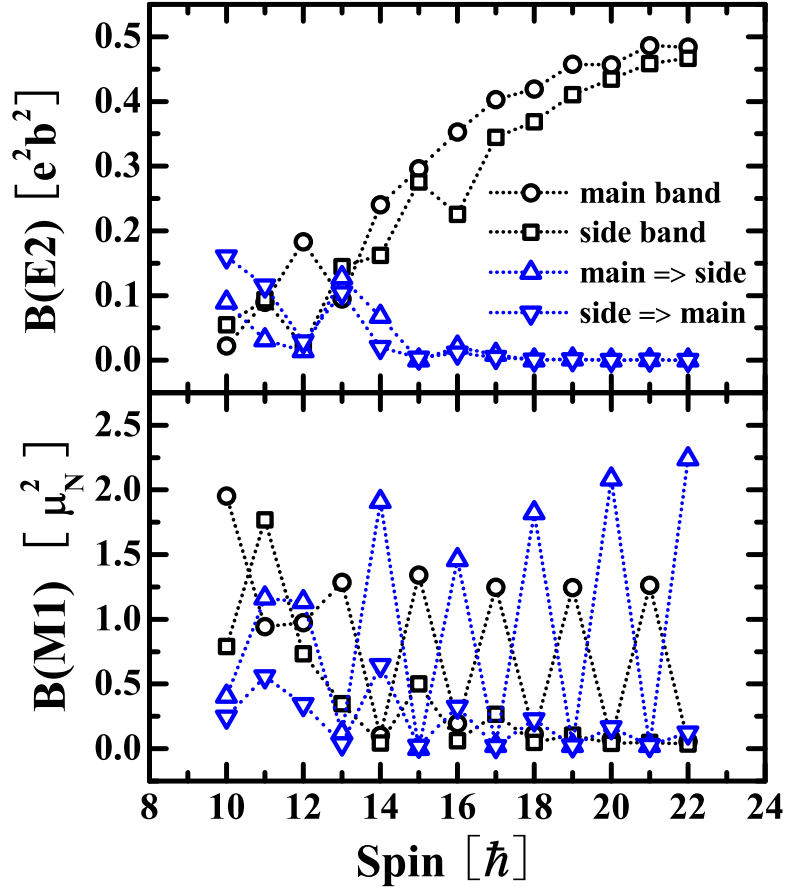}
\caption{\label{fig:BM1andBE2} (Color online) The calculated $B(E2)$
and $B(M1)$ values as functions of the spin in $^{126}$Cs. The
symbols, circles, squares, triangle ups and triangle downs represent
the intraband transitions of the main band, the intraband
transitions of the side band, the interband transitions from the
main band to the side band, and the interband transitions from the
side band to main band, respectively.}
\end{figure}

The lifetime measurement is critical for the identification of
chiral bands. So far, no experimental lifetime data are available
for the candidate chiral bands in $^{126}$Cs. In the present work,
the reduced $B(M1)$ and $B(E2)$ transition probabilities are
calculated and presented in Fig.~\ref{fig:BM1andBE2}. The upper
panel shows the $B(E2)$ transition probabilities, and the lower
panel corresponds to the $B(M1)$ transition probabilities. At low
spins, the intraband and interband $B(E2)$ transition probabilities
are close to each other. After $I=13\hbar$, the intraband $B(E2)$
values steadily increase with spin, whereas interband $B(E2)$ values
steeply decrease and vanish for $I\geq 15\hbar$. It is noted that
both the intraband $B(E2)$ values for the two bands are very
similar, which is consistent with the ideal chiral criteria. The
calculated $B(M1)$ values in the lower panel of the Fig.
\ref{fig:BM1andBE2} show remarkable odd-even staggering. The
intraband $B(M1)$ values at odd spin are larger than the values at
even spin, which has opposite phase with the staggering in the
interband $B(M1)$ values. The staggering amplitude in interband
$B(M1)$ transitions is larger than in the intraband $B(M1)$
transitions. From Fig. \ref{fig:BM1andBE2}, it is clear that the
staggering of the $B(M1)/B(E2)$ ratios is attributed to the
staggering of the $B(M1)$ values.

\begin{figure}
\includegraphics[width=12cm]{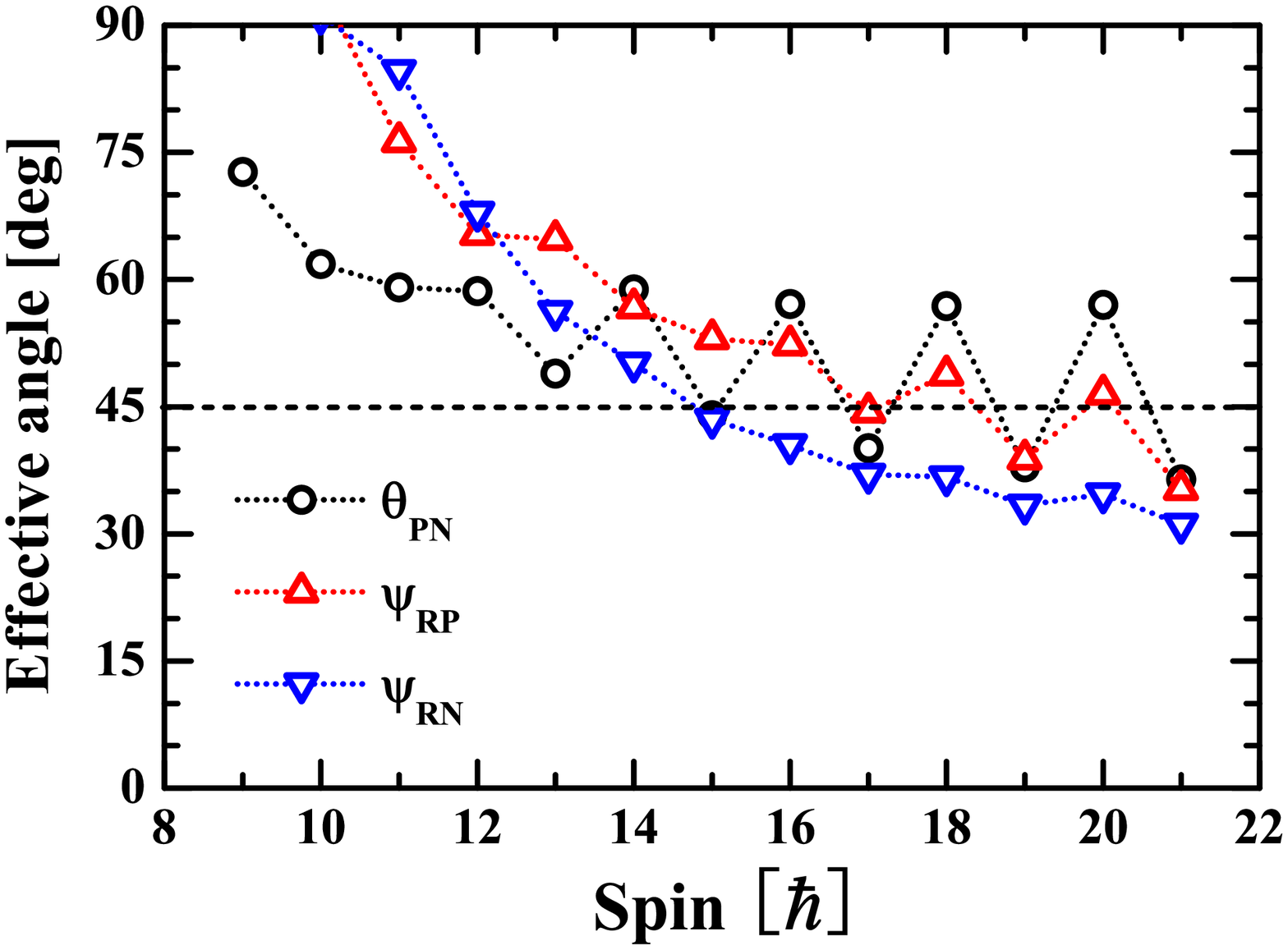}
\caption{\label{fig:anglar} (Color online) The effective angles
$\theta_{\text{PN}}$ (circles), $\psi_{\text{RP}}$ (triangle ups)
and $\psi_{\text{RN}}$ (triangle downs) as a function of spin for
the main band in $^{126}$Cs.}
\end{figure}

To further confirm the picture of the chirality in $^{126}$Cs, the
orientation of the angular momentum for the rotor as well as the
valence proton and neutron are investigated and shown in Fig.
\ref{fig:anglar}. As in Ref.~\cite{Starosta02}, the effective angle
$\theta_{\text{PN}}$ between the angular momenta of the proton
$\textbf{j}_{\textrm{p}}$ and the neutron $\textbf{j}_{\textrm{n}}$
is defined as,
 $\cos \theta_{\text{PN}} =
   \langle \textbf{j}_{\textrm{p}} \cdot \textbf{j}_{\textrm{n}} \rangle /
   \sqrt{\langle j_{\textrm{p}}^2 \rangle \langle j_{\textrm{n}}^{2} \rangle}$.
Similarly, one can define the effective angle $\psi_{\text{RP}}$
($\psi_{\text{RN}}$) between the angular momenta of the core and the
valence proton (neutron). The results of the effective angles are
shown in Fig. \ref{fig:anglar}. In the spin interval 9$\hbar\leq
I\leq$15$\hbar$, the values of three effective angles are larger
than 45$^{\circ}$, which indicates a remarkable aplanar rotation in
this nucleus. This provides additional support for the existence of
chiral bands in $^{126}$Cs.

\section{summary}

In summary, the positive parity doublet bands in $^{126}$Cs based on
the $\pi h_{11/2}\otimes\nu h_{11/2}$ configuration have been
studied by using the two quasi-particles plus a triaxial rotor
model. The energy spectra, energy staggering parameter $S(I)$, the
intraband $B(M1)/B(E2)$ and the
$B(M1)_{\textrm{in}}/B(M1)_{\textrm{out}}$ are calculated and
compared with the available experimental data. After including the
pairing correlation, not only the energy spectra but also the
electromagnetic transition ratios can be well reproduced by the two
quasi-particles coupled with a triaxial rotor model, which supports
the existence of chiral bands in this nucleus. The absolute $B(M1)$
and $B(E2)$ transition probabilities have been presented which is
supposed to invite future experiment on lifetime measurement. The
chiral interpretation of this doublet bands in $^{126}$Cs is also
supported by the effective angles between the angular momenta of the
core, valence proton and neutron. The success of the present two
quasi-particles plus a triaxial rotor model indicates the necessity
for a multi-proton and neutron plus a triaxial rotor model which is
in progress.

\vspace{2em} This work is partly supported by the National Natural
Science Foundation of China under Grant No. 10605001, 10435010,
10221003, and 10505002.

\vspace{1.0cm}
\renewcommand\refname{References}
\renewcommand{\baselinestretch}{1.5}

\newpage

\normalsize

\begin{thebibliography}{100}
\bibitem{FM97} S. Frauendorf and J. Meng, Nucl. Phys. {\bf{A617}}, 131
(1997).
\bibitem{Starosta01} K. Starosta \emph{et al.,} Phys. Rev. Lett. {\bf{86}}, 971
(2001).
\bibitem{Koike01} T. Koike, K. Starosta, C.J. Chiara, D.B. Fossan, and D.R. LaFosse, Phys. Rev. C {\bf{63}}, 061304(R) (2001).
\bibitem{Hartley01} D.J. Hartley \emph{et al.,} Phys. Rev. C {\bf{64}}, 031304(R)
(2001).
\bibitem{Hecht01} A.A. Hecht \emph{et al.,} Phys. Rev. C {\bf{63}}, 051302(R)
(2001).
\bibitem{Bark01} R.A. Bark, A.M. Baxter, A.P. Byrne, G.D. Dracoulis, T. Kibedi, T.R. McGoram, and S.M. Mullins,
 Nucl. Phys. {\bf{A691}}, 577 (2001).
\bibitem{Starosta02} K. Starosta, C. J. Chiara, D. B. Fossan, T. Koike, T. T. S. Kuo, D. R. LaFosse,
 S. G. Rohozinski, Ch. Droste, T. Morek and J. Srebrny Phys. Rev. C {\bf{65}}, 044328
(2002).
\bibitem{Li02} X.F. Li \emph{et al.,} Chin. Phys. Lett. {\bf{19}}, 1779
(2002).
\bibitem{Koike03} T. Koike, K. Starosta, C.J. Chiara, D.B. Fossan, and D.R. LaFosse, Phys. Rev. C {\bf{67}}, 044319
(2003).
\bibitem{GR03} G. Rainovski \emph{et al.,} Phys. Rev. C {\bf{68}}, 024318
(2003)
\bibitem{Vaman04} C. Vaman, D.B. Fossan, T. Koike, K. Starosta, I.Y. Lee and A.O. Macchiavelli, Phys. Rev. Lett. {\bf92}, 032501
(2004).
\bibitem{Joshi04} P. Joshi \emph{et al.,} Phys. Lett. {\bf{B595}}, 135
(2004).
\bibitem{Wang06a} Shouyu Wang, Yunzuo Liu, T. Komatsubara, Yingjun Ma, and Yuhu Zhang, Phys. Rev. C {\bf{74}}, 017302
(2006).
\bibitem{Tonev06} D. Tonev \emph{et al.,} Phys. Rev. Lett. {\bf 96}, 052501 (2006).
\bibitem{Petrache06} C.M. Petrache, G.B. Hagemann, I. Hamamoto and K. Starosta, Phys. Rev. Lett. {\bf{96}}, 112502
(2006).
\bibitem{Grodner06} E. Grodner \emph{et al.}, Phys. Rev. Lett. {\bf97}, 172501 (2006).
\bibitem{Dimitrov00} V.I. Dimitrov, S. Frauendorf, and F. D$\ddot{o}$nau, Phys. Rev. Lett. {\bf{84}}, 5732 (2000).
\bibitem{Frauendorf01} S. Frauendorf, Rev. Mod. Phys. {\bf73}, 463 (2001).
\bibitem{Peng03a} J. Peng, J. Meng, S. Q. Zhang, Phys. Rev. C {\bf{68}}, 044324
(2003).
\bibitem{Peng03b} J. Peng, J. Meng, S. Q. Zhang, Chin. Phys. Lett. {\bf{20}}, 1223
(2003).
\bibitem{Koike04} T. Koike, K. Starosta and I. Hamamoto, Phys. Rev. Lett. {\bf{93}}, 172502
(2004).
\bibitem{Olbratowski04} P. Olbratowski, J. Dobazewski, J. Dudek, and W. P$\l \acute{o}$ciennik, Phys. Rev. Lett. {\bf{93}}, 052501 (2004).
\bibitem{Olbratowski06} P. Olbratowski, J. Dobazewski and J. Dudek, Phys. Rev. C {\bf{73}}, 054308 (2006).
\bibitem{Meyer75} J. Meyer-ter-vehn,  Nucl. Phys. {\bf A249}, 111 (1975).
\bibitem{Ragnarsson88} I. Ragnarsson  and  P.B.~Semmes,  Hyperfine Interaction
{\bf 43}, 425 (1988).
\bibitem{Zhang06} S. Q. Zhang, B. Qi, S. Y. Wang, and J. Meng, to be submitted.
\bibitem{Tajima94} N. Tajima, Nucl. Phys. {\bf{A572}}, 265 (1994).
\bibitem{Fetea05} M.S. Fetea, V. Nikolova and B. Crider, J. Phys. G {\bf{31}}, s1847
(2005).
\bibitem{Koike02} T. Koike \emph{et al.,} FNS2002, Berkeley, CA, 2002, AIP Conf. Proc. No. {\bf 656},
edited by P. Fallon and R. Clark (AIP, Melville, New York, 2003), p.
160.
\end{thebibliography}
\end{document}